\documentclass[
reprent,
preprint,
10pt,        
a4paper,     
twocolumn,   
noshowpacs,  
showkeys,  
amsfonts,    
amsmath,     
amssymb      
]{revtex4-1}

\usepackage[dvipdfmx]{graphicx}
\usepackage{dcolumn}
\usepackage{bm}
\usepackage{times}
\usepackage[usenames]{color}
\usepackage{ulem}

\begin{document}

\title{Thermodynamic quantities of two-dimensional Ising models obtained by noisy mean field annealing and coherent Ising machine}

\author{Kensuke Inaba}
\email{kensuke.inaba.yg@hco.ntt.co.jp}
\author{Yasuhiro Yamada}
\author{Hiroki Takesue}
\affiliation{NTT Basic Research Laboratories, NTT Corporation, 3-1 Morinosato Wakamiya, Atsugi, Kanagawa, 243-0198, Japan}
 
\date{\today}

\begin{abstract} 
Noisy mean field annealing (NMFA) is an algorithm that mimics a coherent Ising machine (CIM), which is an optical system for solving Ising problems. The NMFA has reproduced the solver performance of the CIM  for systems of limited size even though it simplifies the interaction between spins with a mean-field approximation. However, recent experiments observing various thermodynamic quantities have revealed that the CIM can capture the phase transitions of the two-dimensional Ising models that the mean field cannot capture. This situation leads to a fundamental question as to how well the NMFA can capture the features of the thermodynamic quantities around the phase transition. This paper answers that the NMFA reproduces the thermodynamic features of the mean field, but cannot reproduce the CIM results. This suggests that, in terms of sampling, the level of performance of the CIM is beyond that of the NMFA. 
\end{abstract}


\maketitle

\section{Introduction}
An Ising machine is a specific-use computer for solving Ising problems by minimizing the Hamiltonian $H=1/2\sum_{ij} J_{ij} s_{i} s_{j}+\sum_ih_i s_i$, where $s_i(=\pm1)$ is a two-valued Ising spin, $J_{ij}$ is an interaction between spins, and $h_i$ is a magnetic term. Ising problems, which are equivalent to quadratic unconstrained binary optimization problems, can be applied to various problems in the real world \cite{IsingMap,QUBO}. Recent realizations of physical Ising machines using various systems, such  as, trapped ions \cite{IMion}, ultra-cold atoms \cite{IMatom}, superconducting qubits \cite{Dwave,Dwave2}, special CMOS devices \cite{IMcmos}, digital devices \cite{IMfujitsu}, electromechanics \cite{IMelectronic},  magnets \cite{IMmagnet}, and optical systems \cite{OIMLargeScale,OIMFabian,OIMchipbased,OIMsingleshot,DOPOSNN,OIMMF,CIMprinciple,CIMstanford,CIMfirst}, 
have attracted much interest in various research fields. Here, physical principles, e.g., minimizing photon loss in the optical Ising machine \cite{CIMprinciple}, help us to quickly find solutions to Ising problems. These machines have outperformed the conventional algorithms operating on CPUs \cite{CIMfirst,CIMvsDwave,Speed,QuantumSpeedup,SpeedDwave}. On the other hand, algorithms inspired by physical machines encoded with a graphical processing unit (GPU) or a field programmable gate array (FPGA) sometime exhibit better performance than physical machines \cite{NMFA,SimCIM,SB,newSB,PhysicsInspired,CSDE}. However, 
recent progress indicates that physical machines may surpass such algorithms on conventional digital hardware \cite{DwaveNew,CIM10K}. 
Given this state of affairs, we expect that research on computation method will proceed with both physical and physics-inspired solvers.

The coherent Ising machine (CIM) \cite{CIMprinciple,CIMfirst,CIMstanford} is one such optical Ising machine. In the CIM, a degenerate optical parametric oscillator (DOPO) set in a ring cavity is bifurcated by pump pulses to produce signal pulses corresponding to Ising spins \cite{CIMprinciple}. The measurement feedback method \cite{CIMfirst,CIMstanford} is used to implement the interactions between the spins. In this method, parts of the DOPO pulses $c_i$ are extracted from the cavity and measured. The measurements are multiplied by an interaction matrix $J_{ij}$ and are added to the magnetic term $h_i$ by using FPGAs to form feedback. The calculated feedback amplitudes $\phi_i=\sum_jJ_{ij} c_j+h_i$ are injected back to the left portion of the DOPO pulse circulating in the cavity. This procedure to create spin-spin interactions is at first glance mean-field-like. However, the measurement of the coherent optical pulses may cause non-trivial back action on the unmeasured part of DOPO pulses, because the balanced homodyne measurement used here is not a simple projective measurement \cite{Yamada}. Note that the CIM operates near the standard quantum limit with squeezing effects \cite{QuantumLimit}. Thus, the validity of the simple mean-field assumption is suspect if it is applied to the measurement feedback procedure in the CIM. In fact, our recent experiments show that the CIM can capture the phase transition of the two-dimensional Ising model beyond the capacity of the mean-field approximation \cite{2Dpaper}.  Note that another kind of optical Ising machine has reproduced a mean-field description of the phase transition in spin-grass Ising models \cite{OIMMF}.
 
Noisy mean field annealing (NMFA) is a method to solve Ising problems by mimicking a CIM whereby the measurement feedback procedures is simplified to a mean field procedure \cite{NMFA}. The NMFA algorithm is as follows: (i) initialize each spin $s_i$ to be zero; (ii) calculate the feedback $\Phi_i$ with a noise term $\Phi_i=(\sum_j J_{ij} s_j +h_j)/\sqrt{h_i^2+\sum_jJ_{ij}^2}+\xi(0,\sigma)$, where $\xi(0,\sigma)$ is Gaussian noise with standard deviation $\sigma$;
(iii) calculate the mean field equation  $\hat{s}_i=-\tanh(\Phi_i\tilde{\beta})$, where $\tilde{\beta}$ is inverse temperature; 
(iv) replace spin $s_i$ with the convex combination $s_i=\alpha \hat{s}_i+(1-\alpha) s_i$, where $\alpha$ is a feedback constant ($\alpha<1$). 
Procedures (ii)-(iv) are repeated $N_{step}$ times and $\tilde{\beta}$ is increased in each step. 
This approach algorithmically corresponds to a generalized mean field annealing \cite{MFA} with a noise term. In Ref. \cite{NMFA}, King et al. show that the time evolution of $s_i$ in the NMFA gives similar results to those of the optical pulses $c_i$ in the CIM even with a mean-field approximation and stressed that the accuracies of the solutions are comparable to each other. 
 
In this paper, we discuss whether the NMFA is suitable for describing the dynamics of the CIM. For this purpose, we used the NMFA to solve an Ising model problem on a two-dimensional square lattice and then compared those results with the results of a CIM experiment. The Ising model on the square lattice is exactly solvable \cite{2Dexact,2DexactM}, and it is well known that the behaviors of magnetism and other thermodynamic quantities obtained by the mean-field approximation are distinctly different from those exactly obtained. We systematically varied the parameters of the NMFA and revealed that the NMFA and CIM show distinct behaviors in terms of the temperature dependence of the magnetization and other thermodynamic quantities on the square-lattice Ising model. Namely, the CIM captures the features of the exact solution, while the NMFA exhibits the mean-field results. We conclude that the NMFA cannot reach the same level of sampling performance as the CIM, even though its performance in regard to low-energy-state search is comparable to that of the CIM for small systems.

\section{Methods}
Here, we extend the NMFA so it can be used to calculate thermodynamic quantities at a given temperature $T=1/\beta$. First, we rearrange the feedback term as $\Phi_i=\sum_j (J_{ij} s_j +h_j )+\xi(0,\sigma)\sqrt{h_i^2+\sum_jJ_{ij}^2}$. This is because we have to avoid the scale of $\tilde{\beta}$ being renormalized by the term $\sqrt{h_i^2+\sum_jJ_{ij}^2}$. We repeat steps (ii)-(iv) by linearly increasing $\tilde{\beta}$ from 0 to $\beta$ over $N_{\rm step}$ steps. In what follows, we set $N_{\rm step}=10000$. To calculate thermodynamic quantities, we perform sampling and calculate averaged values as follows by repeating the above procedure with $N_{\rm sample}=5000$, where $N_{\rm sample}$ is the number of samples. The internal energy is given by $$U=\frac{1}{N_{\rm sample}} \sum_{k=1}^{N_{\rm sample}}E_k,$$ where $E_k=1/2\sum_{ij} J_{ij} s_i^k s_j^k$ and $s^k_i$ is the $k(=1,\cdots,N_{\rm sample})$-th sampled $i(=1,\cdots,N)$-th spin, where $N$ is the system size. The specific heat is obtained from the statistical equation, 
$$C_{\rm stat}=\beta^2\left(\frac{1}{N_{\rm sample}}\sum_{k=1}^{N_{\rm sample}}E_k^2-U^2\right),$$ and from the thermodynamic relationship, 
$$C_{\rm thermo}=T\frac{\partial S}{\partial T}.$$  The entropy $S$ is calculated from, $$S=\sum_E \bar{P}(E)\ln\frac{\bar{P}(E)}{D(E)},$$ where the energy distribution $\bar{P}(E)$ is obtained from, $\frac{1}{N_{\rm sample}}\sum_k \delta(E-E_k)$, counts of the state with an energy $E$ over $N_{\rm sample}$ in the sampling and $D(E)$ is the density of states of the Ising model explained bellow. The mean root square magnetization is given by $$\sqrt{M^2}=\sqrt{\frac{1}{N_{\rm sample}} \sum_{k=1}^{N_{\rm sample}}(\frac{1}{N}\sum_i s_i^k)^2}.$$ Note that small value of $N_{\rm step}$ and $N_{\rm sample}$ sometimes cause thermodynamic quantities to show unphysical behavior.

Next, we explain the Wang-Landau method \cite{WL} for calculating the ``exact" values of thermodynamic quantities. This method is a kind of Monte Carlo simulation, meaning that it can be only used to a small system, and it allows us to calculate the density of states $D(E)$ of an Ising model, where $E$ is the energy of the model. Upon getting $D(E)$, we can straightforwardly calculate the internal energy, $$U=\sum_E EP(E),$$ the specific heat, $$C_{\rm stat}=\beta^2 \left(\sum_E E^2 P(E)-U^2\right),$$ and the entropy, $$S=-\sum_E P(E)\ln \frac{P(E)}{D(E)},$$ where $P(E)=D(E)\exp(-\beta E)/Z$ and $Z=\sum_E\exp(-\beta E)D(E)$. 
Note that the relation $C_{\rm thermo}=C_{\rm stat}$ can be straightforwardly obtained from the derivative of the above equation of $S$.
To obtain the root mean square magnetization, we simply extend the Wang-Landau method to obtain density of states, $\bar{D}(E,M)$, as a function of magnetization $M=\frac{1}{N} \sum_is_i$  and energy $E$ \cite{WLM}. Here, to obtain $\bar{D}(E,M)$, we perform the Wang-Landau calculation as usual in a limited Hilbert space with a fixed $M$. The mean root square magnetization is given by $$\sqrt{M^2}=\sqrt{ \sum_E\sum_M M^2 \bar{D}(E,M)\exp(-\beta E)/Z}.$$ Within negligibly small numerical errors, the obtained thermodynamic quantities are almost identical to the exact values in a finite size system \cite{WL}.

\section{Results}
\begin{figure}[tb]
\begin{center}
\includegraphics[width=8.6cm]{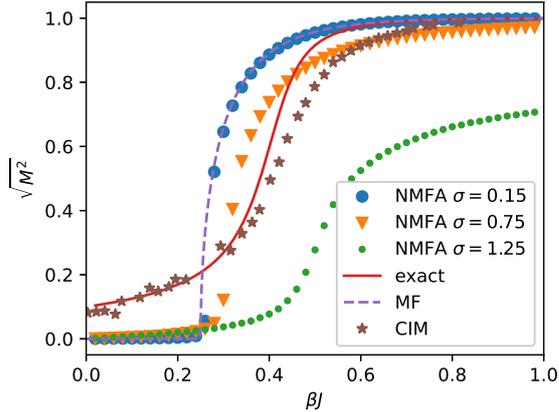}
\end{center}
\caption{Magnetization as a function of $\beta$ on a square $10\times 10$ lattice obtained by the NMFA with $\alpha=0.15$, the Wang Landau method (exact), the mean field approximation (MF), and the CIM.
The CIM results are extracted from Ref. \cite{2Dpaper}.
}
\label{figM}
\end{figure}

First, we revealed that the NMFA reproduces the mean-field behavior in terms of magnetization. In what follows, we consider a two-dimensional $10\times 10$ square lattice with periodic boundary conditions. The Hamiltonian is given by $H=-J/2\sum_{\langle i,j\rangle}s_is_j$, where $\sum_{\langle i,j\rangle}$ indicates a summation over adjacent sites. 
Figure \ref{figM} shows the root mean square magnetization as a function of $\beta J$ for different $\sigma=0.15, 0.75$, and $1.25$ with fixed $\alpha=0.15$. For comparison, Fig. \ref{figM} also shows (noise-less) mean-field results (see Appendix), exact results calculated by the Wang-Landau method \cite{WL}, and experimental results from Ref. \cite{2Dpaper}. Note that, without spontaneous symmetry breaking, the root mean square magnetization has a finite value when the system favors the magnetic ordered phase, and thus, we use this quantity to investigate magnetization in the finite system.  For a small noise deviation $\sigma(=0.15)$, the NMFA results are close to the noiseless mean-field results, while the experimental CIM results are similar to the exact results. This suggests that the measurement feedback of the CIM experiments is not a simple mean-field. For a large $\sigma(=0.75)$, it seems that the NMFA becomes close to the exact results and CIM results, whereas it does not reproduce an important low-temperature behavior, namely, the development of the magnetization up to unity. Namely, the NFMA cannot capture the phase transition to the ferromagnetic state at very low temperatures for large $\sigma$. This is because the large amount of noise behaves like thermal fluctuations. A further increase in $\sigma$ causes the magnetization to show rather unphysical behavior. In addition, we found that, for all $\sigma$, the root mean square magnetization vanishes at high temperatures even though the system size is finite. This is also a characteristic of the mean-field approximation. In contrast, the CIM successfully captures the feature of the magnetization at high temperature.

\begin{figure}[tb]
\begin{center}
\includegraphics[width=8.6cm]{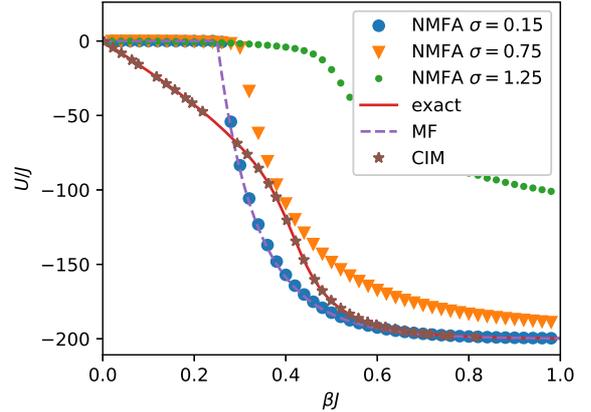}
\end{center}
\caption{Internal energy as a function of $\beta$ on a square $10\times 10$ lattice obtained by the NMFA with $\alpha=0.15$, the Wang Landau method (exact), the mean field approximation (MF), and the CIM.
The CIM results are extracted from Ref. \cite{2Dpaper}.
}
\label{figU}
\end{figure}
\begin{figure}[tb]
\begin{center}
\includegraphics[width=8.6cm]{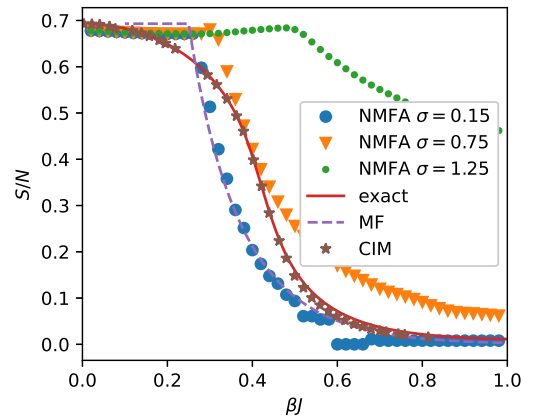}
\end{center}
\caption{Entropy as a function of $\beta$ on a square $10\times 10$ lattice obtained by the NMFA with $\alpha=0.15$, the Wang Landau method (exact), the mean field approximation (MF), and the CIM.
The CIM results are extracted from Ref. \cite{2Dpaper}.
}
\label{figS}
\end{figure}

Figure \ref{figU} and \ref{figS} show the internal energy $U$ and entropy $S$ for the same parameters as those used in Fig. \ref{figM}. In these figures, the NMFA shows mean-field like behavior, such as, $U$ and $S$ plateaus at high temperature and sudden drops from the plateau region with kinks. On the other hand, the CIM captures the features of the exact results, with gradual and smooth changes in both $U$ and $S$. Note that, in the CIM experiments, the inverse temperature $\beta$ was derived from the information on $U$ by using maximally likelihood method, so that the agreement between exact and CIM results is not surprising  (see Ref. \cite{2Dpaper}). However, independently obtained quantity $S$ also shows good agreement, suggesting that the CIM can capture the features of the exact solution. 
It should be noted that $S$ for the NMFA shows rather unphysical behavior whereby $\partial S/\partial\beta>0$ $(\partial S/\partial T<0)$, e.g., around $\beta J=0.7$  with $\sigma=0.15$ and at around $\beta J=0.5$ with $\sigma=1.25$. Namely, the second law of thermodynamics is violated, suggesting the imperfection of the NMFA. For example, an increase in $N_{step}$ does not improve this imperfection.

\begin{figure}[tb]
\begin{center}
\includegraphics[width=8.6cm]{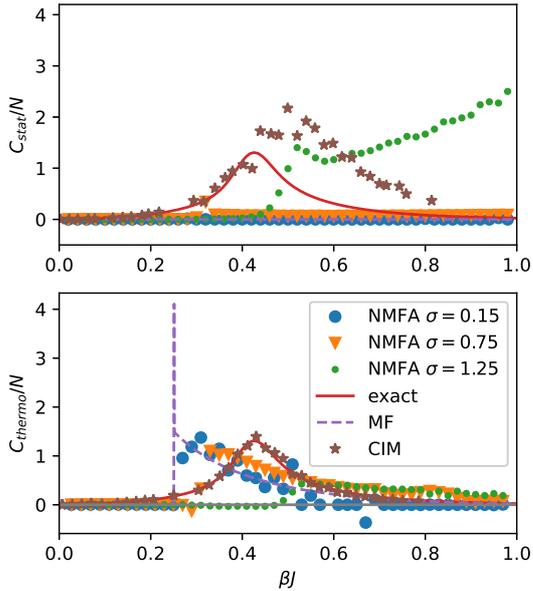}
\end{center}
\caption{Specific heats $C_{\rm stat}$ and $C_{\rm thermo}$ as functions of $\beta$ on a square $10\times 10$ lattice obtained by the NMFA with $\alpha=0.15$, the mean-field approximation (MF), the Wang Landau method (exact), and the CIM.  The CIM results are extracted from Ref. \cite{2Dpaper}.
}
\label{figC}
\end{figure}

Figure \ref{figC} shows specific heats $C_{\rm stat}$ and $C_{\rm thermo}$ for the same parameters as those in Fig. \ref{figM}.  We found that the peak positions of $C_{\rm thermo}$ given by the mean-field approximation are different from the exact results. It is known that the mean-field and the exact solutions in the thermodynamic limit \cite{2Dexact,2DexactM} show different transition points around $\beta J=1/4$ and $\ln(1+\sqrt{2})/2\sim0.44$. Note that, even in the finite system, the Wang-Landau solutions show that the precursor of the transition creates the peak of $C(=C_{\rm thermo}=C_{\rm stat})$ around 0.44. Regarding the CIM results, we found that both $C_{\rm thermo}$ and $C_{\rm stat}$ show clear peak structures around $\beta J\sim0.44$ and $0.5$, respectively, which are closer to the exact solution than that of the mean field. The deviation between these two definitions of specific heat at low temperature indicates that the spins delivered by the CIM are not perfectly sampled from an ideal canonical ensemble. At high temperature, the canonical sampling assumption is satisfied, which is nontrivial and surprising because such a fundamental assumption is not at all assured to be true in the present nonequilibrium open optical system  \cite{2Dpaper}. On the other hand, $C_{\rm stat}$ of the NMFA with $\sigma=0.15$ and 0.75 show flat specific heats, which may be due to the fact that mean-field approximation  neglects some fluctuations (see the Appendix). $C_{\rm stat}$ of the NMFA with $\sigma=1.25$ shows a broad peak at very low temperature, which is different from the exact behavior. The thermodynamic specific heat $C_{\rm thermo}$ of the NMFA is similar to $C_{\rm thermo}$ of the mean-field approximation, while sometimes it takes unphysical negative values, reflecting the violation of the second law of thermodynamics mentioned above.  This unphysical behavior is conspicuous for large $\sigma=1.25$. We should note that the statistical and thermodynamic  relationship $C=\beta^2(\langle H^2\rangle-\langle H\rangle^2)=T \partial S/\partial T$ is not satisfied by the NMFA at all, while such  a relationship is satisfied by the CIM at high temperature \cite{2Dpaper}. Note that the breakdown of the relationship in the CIM results at low temperature is not similar to the breakdown of the NMFA; the CIM captures the qualitative feature, the peak, of the specific heats. This indicates that the CIM reproduces the behavior of specific heats beyond the capacity of a mean-field approximation. 

\begin{figure}[tb]
\begin{center}
\includegraphics[width=8.6cm]{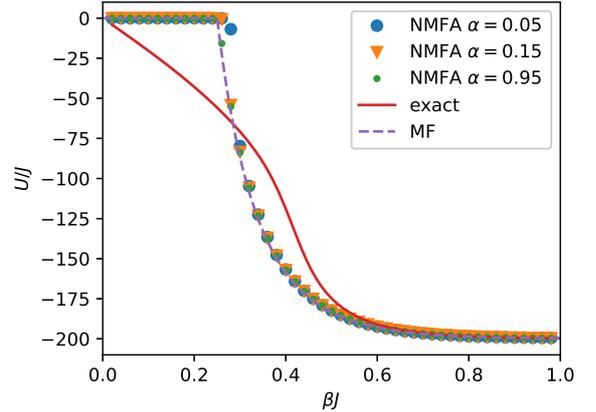}
\end{center}
\caption{Feedback constant $\alpha$ dependence on internal energy of the NMFA with $\sigma=0.15$.}
\label{figalpha}
\end{figure}

Next, we investigated the effect of the parameter $\alpha$. Figure \ref{figalpha} plots $U$ for different $\alpha$ at $\sigma=0.15$. Note that, in the limit $\alpha \to1$, the mean field shown in the Appendix is recovered for small $\sigma$. Thus, $\alpha=0.95$ shows very good agreement with the mean field value. A small $\alpha$ slightly causes a deviation from the mean-field results. However, we found that $\alpha$ does not have effect in the high-temperature region. Other quantities, $S$ and $\sqrt{M^2}$, have similar features (not shown). Thus, the mean-field properties in the high-temperature region, i.e., the plateaus in the quantities and so on as mentioned above, still remain even when $\sigma$ and $\alpha$ are varied. We thus conclude that the NMFA recovers the mean field results and cannot qualitatively reproduce thermodynamic quantities like the CIM can.

\begin{figure}[tb]
\begin{center}
\includegraphics[width=8.6cm]{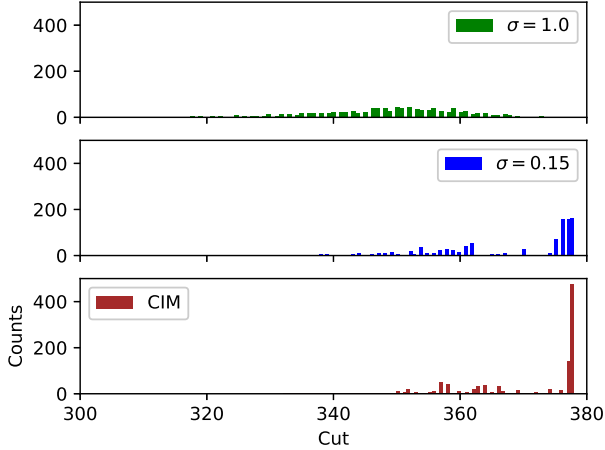}
\end{center}
\caption{Histogram of scores of maximum cut problem, which  takes into account 1000 samples.}
\label{fighist}
\end{figure}

Finally, we discuss how the noise parameter $\sigma$ affects the performance of the solver. For this purpose, we solved a benchmark maximum cut problem with 100 nodes. Figure \ref{fighist} shows a histogram of the scores of this benchmark problem for 1000 trial calculations.  We found that the performance of the solver deteriorate for larger $\sigma$. This tendency coincides with the unphysical behavior at low temperature, where the magnetization does not develop for large $\sigma$ and thus the complete ferromagnetic phase does not appear. In contrast, the CIM  keeps good solver performance when the physical quantities show qualitatively similar behavior to the exact one at low temperatures. Here, the CIM was operated under the same DOPO oscillation conditions as in Ref. \cite{2Dpaper}.

\section{Summary}
In summary, we confirmed that the NMFA reproduces mean-field-like features of various thermodynamic quantities. We also found that specific heat and entropy sometimes show unphysical behaviors, which are due to the mean-field approximation violating the thermodynamic and statistical relationship. In contrast, the CIM captures the exact behaviors of the thermodynamic quantities, in particular, at high temperature \cite{2Dpaper}. At low temperature, the specific heats obtained by the CIM show breakdown of relationship between thermodynamics and statistics. However, the CIM captures the qualitative features of the specific heats; in particular, it reproduces the  qualitative features of the phase transition. This fact suggests that an assumption of statistics, i.e.,  canonical ensembles, will be violated in this region, and a different type of ensemble may reproduce the phase transition. This issue about the breakdown of the statistical assumption, e.g., what kind of ensemble appears, will be left as an important future work for us.
The present results suggest that the measurement feedback of the CIM outperform the mean field approximation, especially, in terms of sampling. The performance of physical Ising machines and algorithms inspired by them as sampler has not been well studied. The future studies on this issue might give us an answer to a lingering question: "what is the killer application of physical machines?"

\appendix 
\section{Mean field}
The mean field approximation starts with the Hamiltonian $H_{MF}=\sum_{ij} J_{ij} s_i \langle s_j\rangle -\sum_{i<j}J_{ij} \langle s_i\rangle\langle s_j\rangle$ with a self-consistent condition $\langle s_i\rangle=-\tanh\beta\Phi_i$ and $\Phi_i=\sum_j J_{ij} \langle s_j\rangle$. $H_{MF}$ is derived from by neglecting the fluctuation term $\langle(s_i-\langle s_i\rangle)(s_j-\langle s_j\rangle )\rangle\sim0$. By considering the translational symmetry, we obtain $\Phi_i=4J\langle s_i\rangle$ for a two-dimensional square lattice. The thermodynamic quantities are as follows: the magnetization $M=\langle s_i\rangle$, the internal energy $U=\langle H_{MF}\rangle=\sum_{i<j} J_{ij} \langle s_i\rangle\langle s_j\rangle=-2NJ\langle s_i\rangle$, the specific heat $C_{\rm thermo}=\partial U/\partial T$, and the entropy $S=\int_0^T C_{\rm thermo}/\bar{T} d\bar{T}$, where derivative and integration in $C_{\rm thermo}$ and $S$ are  done numerically. Note that thermodynamic relationship $C_{\rm thermo}=T\partial S/\partial T$ is surely satisfied.
By using statistical relationships, the  specific heat can be rewritten as $C_{\rm stat}=\beta^2(\langle H_{MF}^2\rangle-\langle H_{MF}\rangle^2 )=4\beta^2\sum_{ij}\sum_{kl}J_{ij} J_{kl} (\langle s_i s_k \rangle-\langle s_i \rangle \langle s_k \rangle)\langle s_j\rangle\langle s_l\rangle $, which reduces to zero. This is due to the fact that the mean field approximation neglect the fluctuation: $\langle s_i s_k\rangle -\langle s_i\rangle\langle s_k \rangle\sim0$. Namely, the mean-field approximation violate a relationship between thermodynamics and statistics, i.e., $C_{\rm thermo}= C_{\rm stat}$. 

\bibliography{NFMA}

\end{document}